\documentclass{article}
\usepackage{ijcai21}
\usepackage{times}
\usepackage{soul}
\usepackage{url}
\usepackage[utf8]{inputenc}
\usepackage[small]{caption}
\usepackage{graphicx}
\usepackage{amsmath}
\usepackage{amsthm}
\usepackage{booktabs}
\usepackage{algorithm}
\usepackage{algorithmic}
\usepackage{graphicx}
\usepackage{comment}
\usepackage{amssymb} 
\usepackage{gensymb}
\usepackage{tikz}
\usepackage{tabularx}

\usepackage{color}

\usepackage[hidelinks,breaklinks]{hyperref}
\usepackage{epsfig}
\usepackage{amstext}
\usepackage{textcomp}
\usepackage{multirow}
\usepackage[T1]{fontenc}

\begin{document}

\title{Medical Image Segmentation Using Squeeze-and-Expansion Transformers}
\author{Shaohua Li\textsuperscript{1}\footnote{Corresponding Author.} \and Xiuchao Sui\textsuperscript{1} \and  Xiangde Luo\textsuperscript{2} \and  Xinxing Xu\textsuperscript{1} \and Yong Liu\textsuperscript{1} \and  Rick Goh\textsuperscript{1} \\
\affiliations
\textsuperscript{1}Institute of High Performance Computing, A*STAR, Singapore\\
\textsuperscript{2}University of Electronic Science and Technology of China, Chengdu, China \\
\emails
\{shaohua, xiuchao.sui\}@gmail.com, xiangde.luo@std.uestc.edu.cn,
\{xuxinx, liuyong, gohsm\}@ihpc.a-star.edu.sg
}

\maketitle
\begin{abstract}
Medical image segmentation is important for computer-aided diagnosis. Good segmentation demands the model to see the big picture and fine details simultaneously, i.e., to learn image features that incorporate large context while keep high spatial resolutions. To approach this goal, the most widely used methods -- U-Net and variants, extract and fuse multi-scale features. However, the fused features still have small \emph{effective receptive fields} with a focus on local image cues, limiting their performance. In this work, we propose Segtran, an alternative segmentation framework based on transformers, which have unlimited \emph{effective receptive fields} even at high feature resolutions. The core of Segtran is a novel Squeeze-and-Expansion transformer: a squeezed attention block regularizes the self attention of transformers, and an expansion block learns diversified representations. Additionally, we propose a new positional encoding scheme for transformers, imposing a continuity inductive bias for images. Experiments were performed on 2D and 3D medical image segmentation tasks: optic disc/cup segmentation in fundus images (REFUGE'20 challenge), polyp segmentation in colonoscopy images, and brain tumor segmentation in MRI scans (BraTS'19 challenge). Compared with representative existing methods, Segtran consistently achieved the highest segmentation accuracy, and exhibited good cross-domain generalization capabilities. The source code of Segtran is released at \url{https://github.com/askerlee/segtran}.
\end{abstract}

\section{Introduction}
\begin{figure*}
\centering
  \includegraphics[scale=0.4]{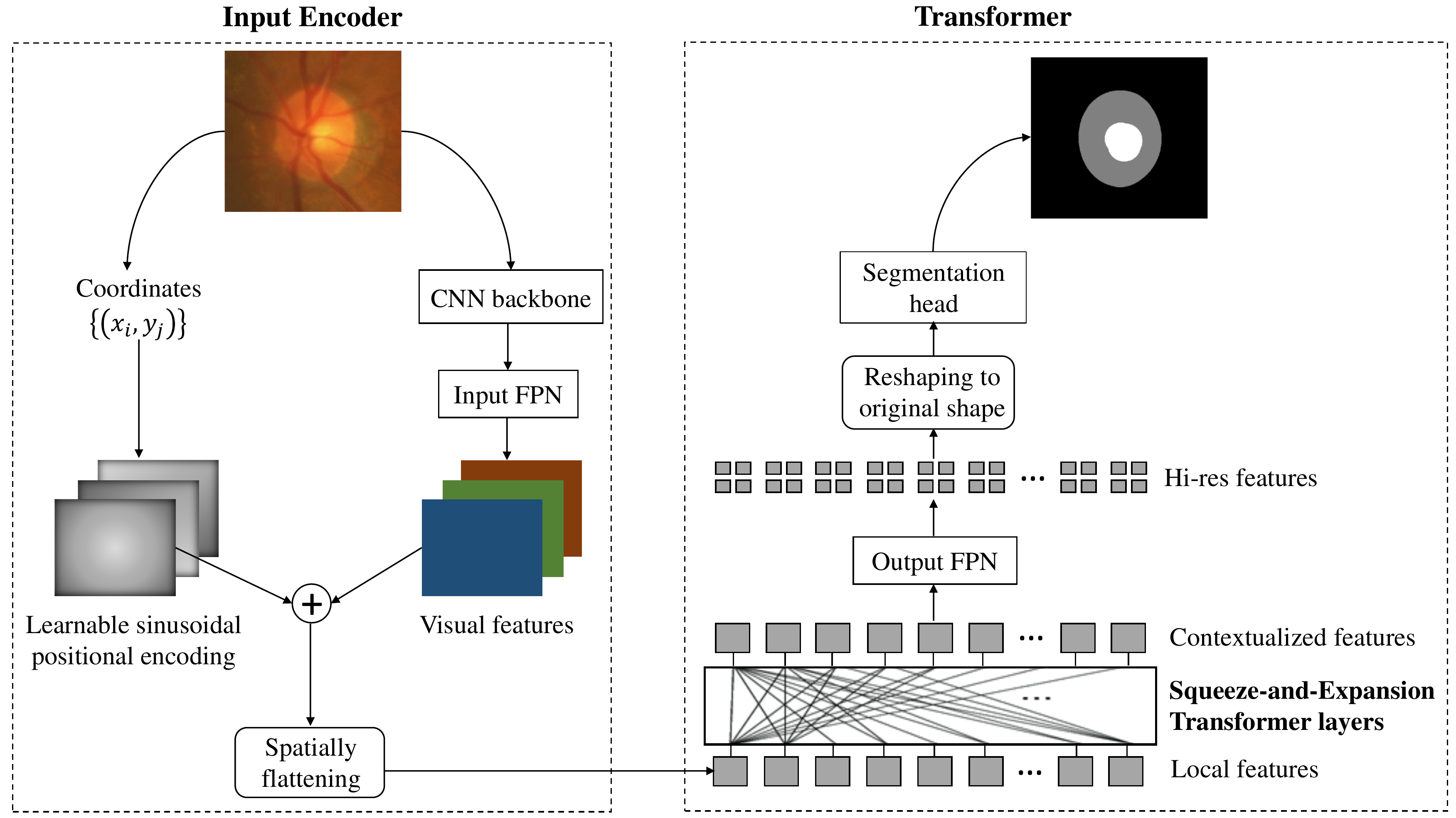}
  \captionof{figure}{Segtran architecture. It extracts visual features with a CNN backbone, combines them with positional encodings of the pixel coordinates, and flattens them into a sequence of local feature vectors. The local features are contextualized by a few Squeeze-and-Expansion transformer layers. To increase spatial resolution, an input FPN and an output FPN upsamples the features before and after the transformers.}
  \label{fig:segtran-arch}
\end{figure*}

Automated Medical image segmentation, i.e., automated delineation of anatomical structures and other regions of interest (ROIs), is an important step in computer-aided diagnosis; for example it is used to quantify tissue volumes, extract key quantitative measurements, and localize pathology \cite{attention-unet-journal,refuge}. Good segmentation demands the model to see the big picture and fine details at the same time, i.e., learn image features that incorporate large context while keep high spatial resolutions to output fine-grained segmentation masks. However, these two demands pose a dilemma for CNNs, as CNNs often incorporate larger context at the cost of reduced feature resolution. A good measure of how large a model ``sees'' is the \emph{effective receptive field} (effective RF) \cite{eff-rf}, i.e., the input areas which have non-negligible impacts to the model output.

\begin{figure}[t]
\centering
  \includegraphics[scale=0.25]{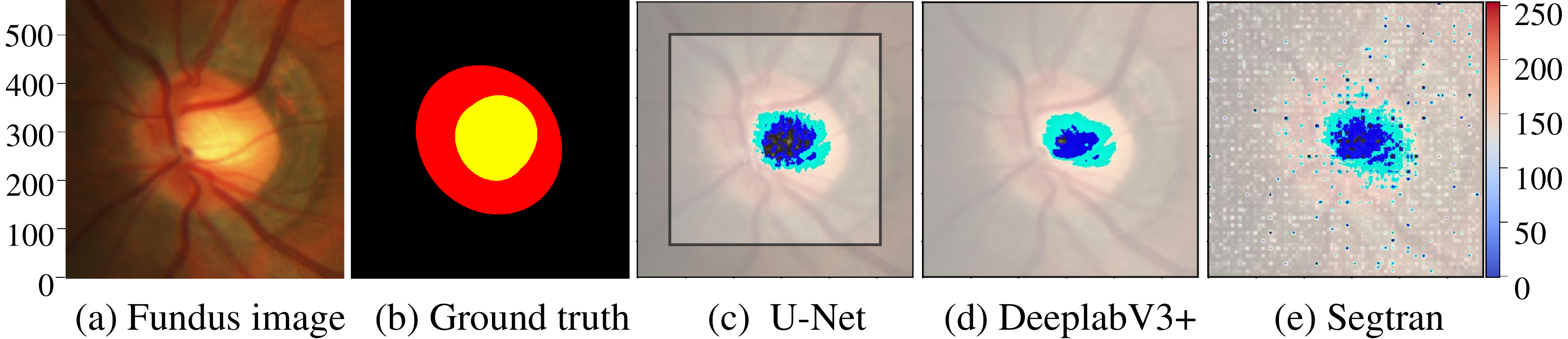}
  \captionof{figure}{Effective receptive fields of 3 models, indicated by non-negligible gradients in blue blobs and light-colored dots. Gradients are back-propagated from the center of the image. Segtran has non-negligible gradients dispersed across the whole image (light-colored dots). U-Net and DeepLabV3+ have concentrated gradients. Input image: $576\times 576$.}
  \label{fig:eff-RF}
\end{figure}

Since the advent of U-Net \cite{unet}, it has shown excellent performance across medical image segmentation tasks. A U-Net consists of an encoder and a decoder, in which the encoder progressively downsamples the features and generates coarse contextual features that focus on contextual patterns, and the decoder progressively upsamples the contextual features and fuses them with fine-grained local visual features. The integration of multiple scale features enlarges the RF of U-Net, accounting for its good performance. However, as the convolutional layers deepen, the impact from far-away pixels decay quickly. As a results, the effective RF of a U-Net is much smaller than its theoretical RF. As shown in Fig.\ref{fig:eff-RF}, the effective RFs of a standard U-Net and DeepLabV3+ are merely around 90 pixels. This implies that they make decisions mainly based on individual small patches, and have difficulty to model larger context. However, in many tasks, the heights/widths of the ROIs are greater than 200 pixels, far beyond their effective RFs.  Without ``seeing the bigger picture'', U-Net and other models may be misled by local visual cues and make segmentation errors. 

Many improvements of U-Net have been proposed. A few typical variants include: U-Net++ \cite{unet++} and U-Net 3+ \cite{unet3+}, in which more complicated skip connections are added to better utilize multi-scale contextual features; attention U-Net \cite{attention-unet-journal}, which employs attention gates to focus on foreground regions; 3D U-Net \cite{3d-unet} and V-Net \cite{vnet}, which extend U-Net to 3D images, such as MRI volumes; Eff-UNet \cite{eff-unet}, which replaces the encoder of U-Net with a pretrained EfficientNet \cite{efficientnet}.

Transformers \cite{transformer} are increasingly popular in computer vision tasks. A transformer calculates the pairwise interactions (``self-attention'') between all input units, combines their features and generates contextualized features. The contextualization brought by a transformer is analogous to the upsampling path in a U-Net, except that it has unlimited effective receptive field, good at capturing long-range correlations. 
Thus, it is natural to adopt transformers for image segmentation. In this work, we present Segtran, an alternative segmentation architecture based on transformers. A straightforward incorporation of transformers into segmentation only yields moderate performance. As transformers were originally designed for Natural Language Processing (NLP) tasks, there are several aspects that could be improved to better suit image applications. To this end, we propose a novel transformer design \emph{Squeeze-and-Expansion Transformer}, in which a squeezed attention block helps regularize the huge attention matrix, and an expansion block learns diversified representations. In addition, we propose a  learnable sinusoidal positional encoding that imposes a continuity inductive bias for the transformer. Experiments demonstrate that they lead to improved segmentation performance.


We evaluated Segtran on two 2D medical image segmentation tasks: optic disc/cup segmentation in fundus images of the REFUGE'20 challenge, and polyp segmentation in colonoscopy images. Additionally, we also evaluated it on a 3D image segmentation task:  brain tumor segmentation in MRI scans of the BraTS'19 challenge. Segtran has consistently shown better performance than U-Net and its variants (UNet++, UNet3+, PraNet, and nnU-Net), as well as DeepLabV3+ \cite{deeplabv3+}. 

\section{Related Work}
Our work is largely inspired by DETR \cite{detr}. DETR uses transformer layers to generate contextualized features that represent objects, and learns a set of object queries to extract the positions and classes of objects in an image. Although DETR is also explored to do panoptic segmentation \cite{panoptic}, it adopts a two-stage approach which is not applicable to medical image segmentation. A followup work of DETR, Cell-DETR \cite{cell-detr} also employs transformer for biomedical image segmentation, but its architecture is just a simplified DETR, lacking novel components like our Squeeze-and-Expansion transformer. Most recently, SETR \cite{SETR} and TransU-Net \cite{transunet} were released concurrently or after our paper submission. Both of them employ a Vision Transformer (ViT) \cite{vision-trans} as the encoder to extract image features, which already contain global contextual information. A few convolutional layers are used as the decoder to generate the segmentation mask. In contrast, in Segtran, the transformer layers build global context on top of the local image features extracted from a CNN backbone, and a Feature Pyramid Network (FPN) generates the segmentation mask.

\cite{add-pos-to-cnn} extends CNNs with positional encoding channels, and evaluates them on segmentation tasks. Mixed results were observed. In contrast, we verified through an ablation study that positional encodings indeed help Segtran to do segmentation to a certain degree.

Receptive fields of U-Nets may be enlarged by adding more downsampling layers. However, this increases the number of parameters and adds the risk of overfitting. Another way of increasing receptive fields is using larger stride sizes of the convolutions in the downsampling path. Doing so, however, sacrifices spatial precision of feature maps, which is often disadvantageous for segmentation \cite{coarse-features}.


\section{Squeeze-and-Expansion Transformer} \label{s-e-trans}
The core concept in a transformer is \emph{Self Attention}, which can be understood as computing an affinity matrix between different units, and using it to aggregate features:
\begin{align}
\operatorname{Att\_weight}(\boldsymbol{X},\boldsymbol{X}) & = f(\boldsymbol{K}(\boldsymbol{X}), \boldsymbol{Q}(\boldsymbol{X})) \in \mathbb{R}^{N \times N}, \\
 \operatorname{Attention}(\boldsymbol{X}) &= \operatorname{Att\_weight}(\boldsymbol{X},\boldsymbol{X}) \cdot \boldsymbol{V}(\boldsymbol{X}), \label{attention-eqn} \\ 
\boldsymbol{X}_{out} &= \operatorname{FFN}(\operatorname{Attention}(\boldsymbol{X})), 
\end{align}
where $\boldsymbol{K}, \boldsymbol{Q}, \boldsymbol{V}$ are key, query, and value projections, respectively. $f$ is softmax after dot product. $\operatorname{Att\_weight}(\boldsymbol{X},\boldsymbol{X})$ is the pairwise attention matrix between input units, whose $i,j$-th element defines how much the features of unit $j$ contributes to the fused (contextualized) features of unit $i$.  $\operatorname{FFN}$ is a feed-forward network to further transform the fused features.

The basic transformer above is extended to a \emph{multi-head attention} (MHA) \cite{transformer,analyze-multi-head}, aiming to capture different types of associations between input units. Each of the $N_h$ heads computes individual attention wights and output features ($C/N_h$-dimensional), and their output features are \emph{concatenated} along the channel dimension into $C$-dimensional features. Different heads operate in exclusively different feature subspaces. 

We argue that transformers can be improved in four aspects make them better suited for images:
\begin{enumerate}
\item In Eq.\eqref{attention-eqn}, the intermediate features $\operatorname{Attention}(\boldsymbol{X})$ are obtained by linearly combining the projected input features, where the attention matrix specifies the combination coefficients. As the attention matrix is huge: $N\times N$, with typically $N > 1000$, it is inherently vulnerable to noises and  overfitting. Reducing the attention matrix to lower rank matrices may help.
\item In traditional transformers, the output features are \emph{monomorphic}: it has only one set of feature transformations (the multi-head transformer also has \emph{one set} of transformations after concatenation), which may not have enough capacity to fully model data variations. Just like a mixture of Gaussians almost always better depicts a data distribution than a single Gaussian, data variations can be better captured using a mixture of $k$ transformers.
\item In traditional transformers, the key and query projections are independently learned, enabling them to capture \emph{asymmetric} relationships between tokens in natural language. However, the relationships between image units are often \emph{symmetric}, such as whether two pixels belong to the same segmentation class.
\item Pixels in an image have strong locality and semantic continuity. The two mainstream positional encoding schemes \cite{detr,vision-trans} do not fully impose such an \emph{inductive bias}. This bias could be imposed by an improved positional encoding.
\end{enumerate}

The Squeeze-and-Expansion Transformer aims to improve in all the four aspects. The \emph{Squeezed Attention Block} computes attention between the input and $M$ inducing points \cite{set-trans}, and compresses the attention matrices to $N\times M$. The \emph{Expanded Attention Block} is a mixture-of-experts model with $N_m$ modes (``experts''). In both blocks, the query projections and key projections are \emph{tied} to make the attention \emph{symmetric}, for better modeling of the symmetric relationships between image units. In addition, a \emph{Learnable Sinusoidal Positional Encoding} helps the model capture spatial relationships.

\begin{figure}[t]
\centering
  \includegraphics[scale=0.35]{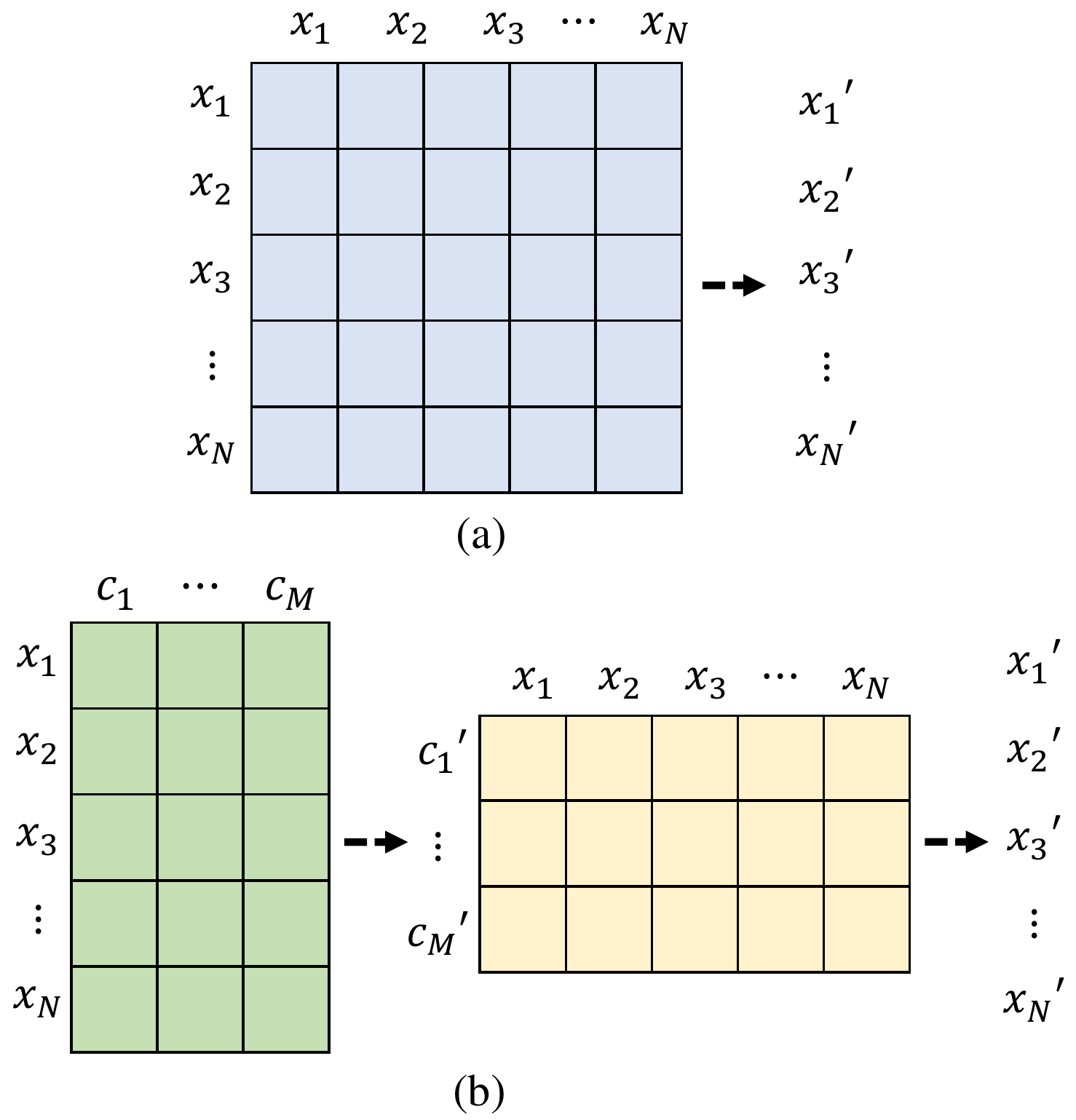}
  \captionof{figure}{(a) Full self-attention ($N\times N$) vs. (b) Squeezed Attention Block (SAB). In SAB, first input units $\boldsymbol{x}_1,\cdots,\boldsymbol{x}_N$  attend with a codebook $\boldsymbol{c}_1,\cdots,\boldsymbol{c}_M$, yielding projected codebook features $\boldsymbol{c}_1',\cdots,\boldsymbol{c}_M'$, which then attend back with the input $\boldsymbol{x}_1,\cdots,\boldsymbol{x}_N$. The two attention matrices are $N\times M$ and $M\times N$, respectively.}
  \label{fig:sab}
\end{figure}

\begin{figure}[t]
\centering
  \includegraphics[scale=0.35]{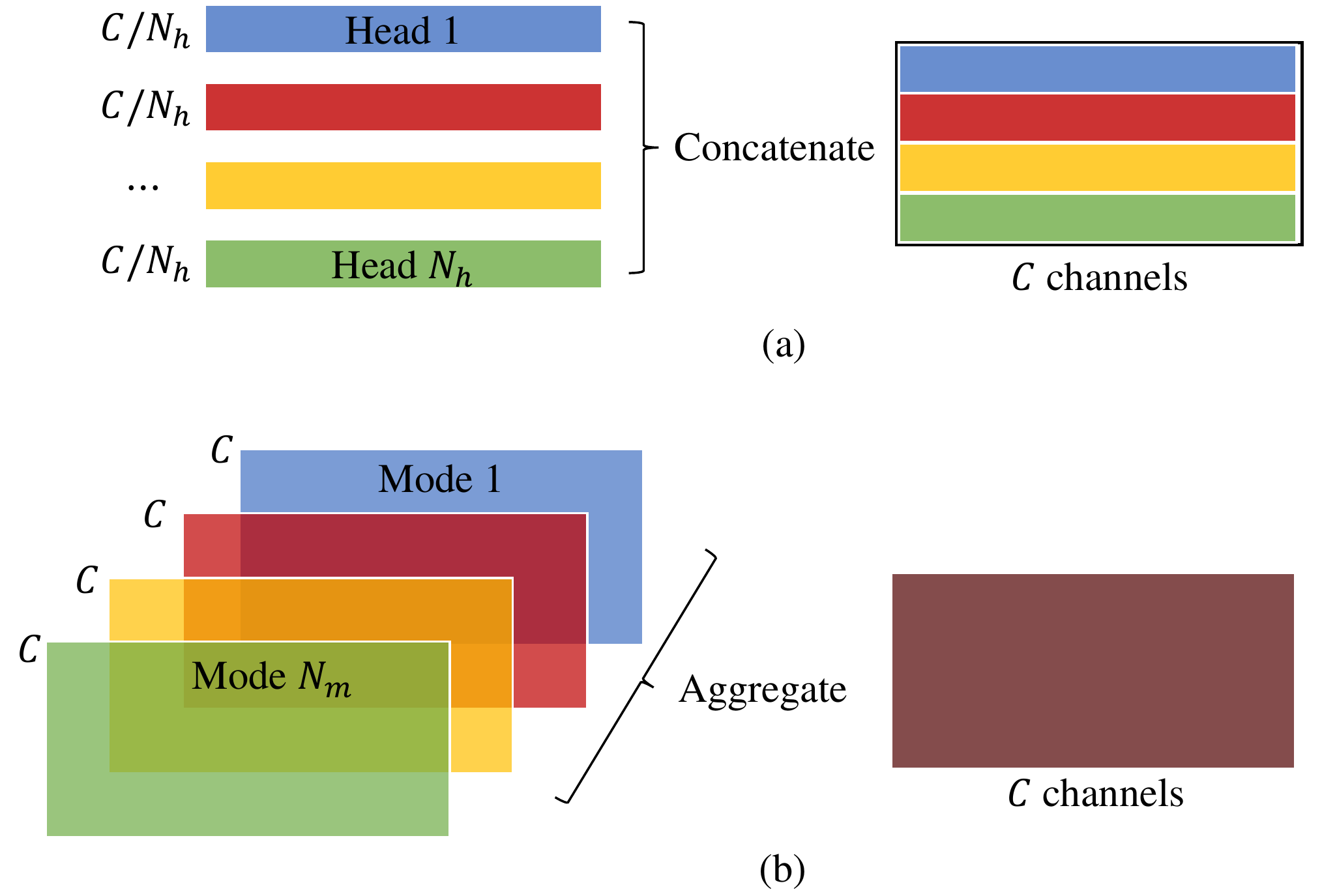}
  \captionof{figure}{(a) Multi-head attention (MHA) vs. (b) Expanded attention block (EAB). In MHA, each head outputs an exclusive feature subset. In contrast, EAB outputs $N_m$ sets of complete features from $N_m$ modes, and aggregates them with dynamic mode attention.}
  \label{fig:eab}
\end{figure}

\subsection{Squeezed Attention Block}
\cite{set-trans} proposes \emph{Induced Set Attention Block} (ISAB) by bringing \emph{inducing points} into the transformer. It was originally designed to learn good features of a set of unordered objects. Here we employ this design to ``squeeze'' the bloated attention matrix, so as to reduce noises and overfitting in image tasks. We rename ISAB as \emph{Squeezed Attention Block} (SAB) to highlight its new role in this context\footnote{We clarify that our contribution is a novel transformer architecture that combines SAB with an Expanded Attention Block.}.

In SAB, inducing points are a set of $M$ learned embeddings $\boldsymbol{c}_1,\cdots,\boldsymbol{c}_M$ in an external \emph{discrete codebook}. Usually $M \ll N$, the number of input units. The inducing points are first transformed into new embeddings $\boldsymbol{C}' = \boldsymbol{c}_1',\cdots,\boldsymbol{c}_M'$ after attending with the input. The combination of these embeddings form the output features $\boldsymbol{X}_{out} = \boldsymbol{x}_1',\cdots,\boldsymbol{x}_N'$ (Fig.\ref{fig:sab}):
\begin{align}
\boldsymbol{C}' &= \text{Single-Head}(\boldsymbol{X}, \boldsymbol{C}), \\
\boldsymbol{X}_{out} &= \text{EAB}(\boldsymbol{C}', \boldsymbol{X}),
\end{align}
where $\text{Single-Head}(\cdot,\cdot)$ is a single-head transformer, and $\text{EAB}(\cdot,\cdot)$ is an Expanded Attention Block presented in the next subsection.
In each of the two steps, the attention matrix is of $N\times M$, much more compact than vanilla transformers. 

SAB is conceptually similar to the codebook used for discrete representation learning in \cite{taming}, but 
the discretized features are further processed by a transformer.
SAB can trace its lineage back to low-rank matrix factorization, i.e., approximating a data matrix $\boldsymbol{X}_{n\times n}\approx \boldsymbol{P}_{n\times d} \cdot \boldsymbol{Q}_{d\times n}$, which is a traditional regularization technique against data noises and overfitting. Confirmed by an ablation study, SAB helps fight against noises and overfitting as well.

\subsection{Expanded Attention Block}
The Expanded Attention Block (EAB) consists of $N_m$ modes, each an individual single-head transformer. They output $N_m$ sets of contextualized features, which are then aggregated into one set using dynamic mode attention:
\begin{align}
\boldsymbol{X}_{out}^{(k)} =& \operatorname{Mode}^{(k)}(\boldsymbol{X}), \\
\boldsymbol{B}^{(k)} =& \operatorname{Linear}^{(k)}(\boldsymbol{X}_{out}^{(k)}), \label{mode-attn} \\
\text{with } &k \in \{1,\cdots,N_m\}, \nonumber \\
\boldsymbol{G} =& \operatorname{softmax}\left(\boldsymbol{B}^{(1)}, \cdots, \boldsymbol{B}^{(N_m)}\right), \label{mode-attn-softmax} \\
\boldsymbol{X}_{out} =& \left(\boldsymbol{X}_{out}^{(1)}, \cdots, \boldsymbol{X}_{out}^{(N_m)} \right) \cdot \boldsymbol{G}^\top, \label{mode-avg}
\end{align}
where the mode attention $G \in \mathbb{R}^{N_u \times N_m}$ is obtained by doing a linear transformation of each mode features, and taking softmax over all the modes. Eq.\eqref{mode-avg} takes a weighted sum over the modes to get the final output features $\boldsymbol{X}_{out}$. This dynamic attention is inspired by the Split Attention of the ResNest model \cite{resnest}.

EAB is a type of Mixture-of-Experts \cite{sparse-moe}, an effective way to increase model capacity. Although there is resemblance between multi-head attention (MHA) and EAB, they are essentially different, as shown in Fig.\ref{fig:eab}. In MHA, each head resides in an exclusive feature subspace and provides unique features. In contrast, different modes in EAB share the same feature space, and the representation power largely remains after removing any single mode. The modes join forces to offer more capacity to model diverse data, as shown in an ablation study. In addition, EAB is also different from the  Mixture of Softmaxes (MoS) transformer \cite{fpt}. Although MoS transformer also uses $k$ sets of queries and keys, it shares one set of value transformation.

\subsection{Learnable Sinusoidal Positional Encoding}
A crucial inductive bias for images is the pixel locality and semantic continuity, which is naturally encoded by convolutional kernels. As the input to transformers is flattened into 1-D sequences, positional encoding (PE) is the only source to inject information about spatial relationships. On the one hand, this makes transformers flexible to model arbitrary shapes of input. On the other hand, the continuity bias of images is non-trivial to fully incorporate. This is a limitation of the two mainstream PE schemes: Fixed Sinusoidal Encoding and Discretely Learned Encoding \cite{detr,vision-trans}. The former is spatially continuous but lacks adaptability, as the code is predefined. The latter learns a discrete code for each coordinate without enforcing spatial continuity.

We propose \emph{Learnable Sinusoidal Positional Encoding}, aiming to bring in the continuity bias with adaptability. 
Given a pixel coordinate $(x,y)$, our positional encoding vector $\boldsymbol{pos}(x,y)$ is a combination of sine and cosine functions of linear transformations of $(x,y)$:
\begin{equation}
pos_i(x,y) = \begin{cases}
\sin(a_i x + b_i y + c_i) \quad \text{if } i < C/2 \\
\cos(a_i x + b_i y + c_i) \quad \text{if } i \ge C/2, 
\end{cases}\label{pos-enc}
\end{equation}
where $i$ indexes the elements in $\boldsymbol{pos}$, $\{a_i,b_i,c_i\}$ are learnable weights of a linear layer, and $C$ is the dimensionality of image features. To make the PE behave consistently across different image sizes, we normalize $(x,y)$ into $[0, 1]^2$. When the input image is 3D, Eq.\eqref{pos-enc} is trivially extended by using 3D coordinates $(x,y,z)$.

The encoding in Eq.\eqref{pos-enc} changes smoothly with pixel coordinates, and thus nearby units receive similar positional encodings, pushing the attention weights between them towards larger values, which is the spirit of the continuity bias. The learnable weights and sinusoidal activation functions make the code both adaptable and nonlinear to model complex spatial relationships \cite{fourier}. 

\section{Segtran Architecture}
As a context-dependent pixel-wise classification task, segmentation faces a conflict between larger context (lower resolution) and localization accuracy (higher resolution). Segtran partly circumvents this conflict by doing pairwise feature contextualization, without sacrificing spatial resolutions. 
There are five main components in Segtran (Fig.\ref{fig:segtran-arch}): 1) a CNN backbone to extract image features, 2) input/output feature pyramids to do upsampling, 3) learnable sinusoidal positional encoding, 4) Squeeze-and-Expansion transformer layers to contextualize features, and 5) a segmentation head.

\subsection{CNN Backbone} \label{backbone}
We employ a pretrained CNN backbone to extract features maps with rich semantics. Suppose the input image is $X_0\in \mathbb{R}^{H_0 \times W_ 0\times D_0}$,  where for a 2D image, $D_0=1$ or 3 is the number of color channels. For a 3D image, $D_0\gg 3$ is the number of slices in the depth dimension. For 2D and 3D images, the extracted features are $\text{CNN}(X_0)\in \mathbb{R}^{C\times H \times W}$, and $\text{CNN}(X_0)\in \mathbb{R}^{C\times H \times W \times D}$, respectively.

On 2D images, typically ResNet-101 or EfficientNet-D4 is used as the backbone. For increased spatial resolution, we change the stride of the first convolution from 2 to 1. Then $H,W=H_0/16,W_0/16$. On 3D images, 3D backbones like I3D \cite{i3d} could be adopted.

\subsection{Transformer Layers}
Before being fed into the transformer, the visual features and positional encodings of each unit are added up before being fed to the transformer: $\boldsymbol{X}_{\text{spatial}}=\boldsymbol{X}_{\text{visual}} + \boldsymbol{pos}(\text{coordinates}(\boldsymbol{X}))$. 
$\boldsymbol{X}_{\text{spatial}}$ is flattened across spatial dimensions to a 1-D sequence $\boldsymbol{X}_0 \in \mathbb{R}^{N_u \times C}$, where $N_u$ is the total number of image units, i.e., points in the feature maps.

The transformer consists of a few stacked transformer layers. Each layer takes input features $\boldsymbol{X}$, computes the pairwise interactions between input units, and outputs contextualized features $\boldsymbol{X}_{out}$ of the same number of units.
The transformer layers used are our novel design \emph{Squeeze-and-Expansion Transformer} (Section \ref{s-e-trans}).

\subsection{Feature Pyramids and Segmentation Head}
Although the spatial resolution of features is not reduced after passing through the transformer layers, for richer semantics, the input features to transformers are usually high-level features from the backbone. They are of a low spatial resolution, however. Hence, we increase their spatial resolution with an input Feature Pyramid Network (FPN) \cite{panet} and an output FPN, which upsample the feature maps at the transformer input end and output end, respectively.

Without loss of generality, let us assume the EfficientNet is the backbone. The stages 3, 4, 6, and 9 of the network are commonly used to extract multi-scale feature maps. Let us denote the corresponding feature maps as $\boldsymbol{f}_1, \boldsymbol{f}_2, \boldsymbol{f}_3, \boldsymbol{f}_4$, respectively. Their shapes are $\boldsymbol{f}_i\in \mathbb{R}^{C_i \times H_i \times W_i}$, with $H_i = \frac{H_0}{2^i}, W_i = \frac{W_0}{2^i}$.

As described above, $\boldsymbol{f}(X_0) = \boldsymbol{f}_4$ is $1/16$ of the original image, which is too coarse for accurate segmentation. Hence, we upsample it with an \textbf{input FPN}, and obtain upsampled feature maps $\boldsymbol{f}_{34}$:
\begin{equation}
    \boldsymbol{f}_{34} = \operatorname{upsample}_{\times 2}(\boldsymbol{f}_4) + \operatorname{conv}_{34}(\boldsymbol{f}_3),
\end{equation}
where $\operatorname{conv}_{34}$ is a $1\times 1$ convolution that aligns the channels of $\boldsymbol{f}_3$ to $\boldsymbol{f}_4$, and $\operatorname{upsample}_{\times 2}(\cdot)$ is bilinear interpolation.

$\boldsymbol{f}_{34}$ is $1/8$ of the original image, and is used as the input features to the transformer layers. As the transformer layers keep the spatial resolutions unchanged from input to output feature maps, the output feature maps $\boldsymbol{g}_{34}$ is also $1/8$ of the input image. Still, this spatial resolution is too low for segmentation. Therefore, we adopt an \textbf{output FPN} to upsample the feature maps by a factor of 4 (i.e., $1/2$ of the original images). The output FPN consists of two upsampling steps:
\begin{align}
    \boldsymbol{f}_{12} &= \operatorname{upsample}_{\times 2}(\boldsymbol{f}_2) + \operatorname{conv}_{12}(\boldsymbol{f}_1), \nonumber \\
    \boldsymbol{g}_{1234} &= \operatorname{upsample}_{\times 4}(\boldsymbol{g}_{34}) + \operatorname{conv}_{24}(\boldsymbol{f}_{12}),
\end{align}
where $\operatorname{conv}_{12}$ and $\operatorname{conv}_{24}$ are $1\times 1$ convolutional layers that align the channels of $\boldsymbol{f}_1$ to $\boldsymbol{f}_2$, and $\boldsymbol{f}_2$ to $\boldsymbol{f}_4$, respectively.

This FPN scheme is the bottom-up FPN proposed in \cite{panet}. Empirically, it performs better than the original top-down FPN \cite{fpn}, as richer semantics in top layers are better preserved.

The segmentation head is simply a $1\times 1$ convolutional layer, outputting confidence scores of each class in the mask.

\section{Experiments}
\begin{figure}
\centering
  \includegraphics[scale=0.3]{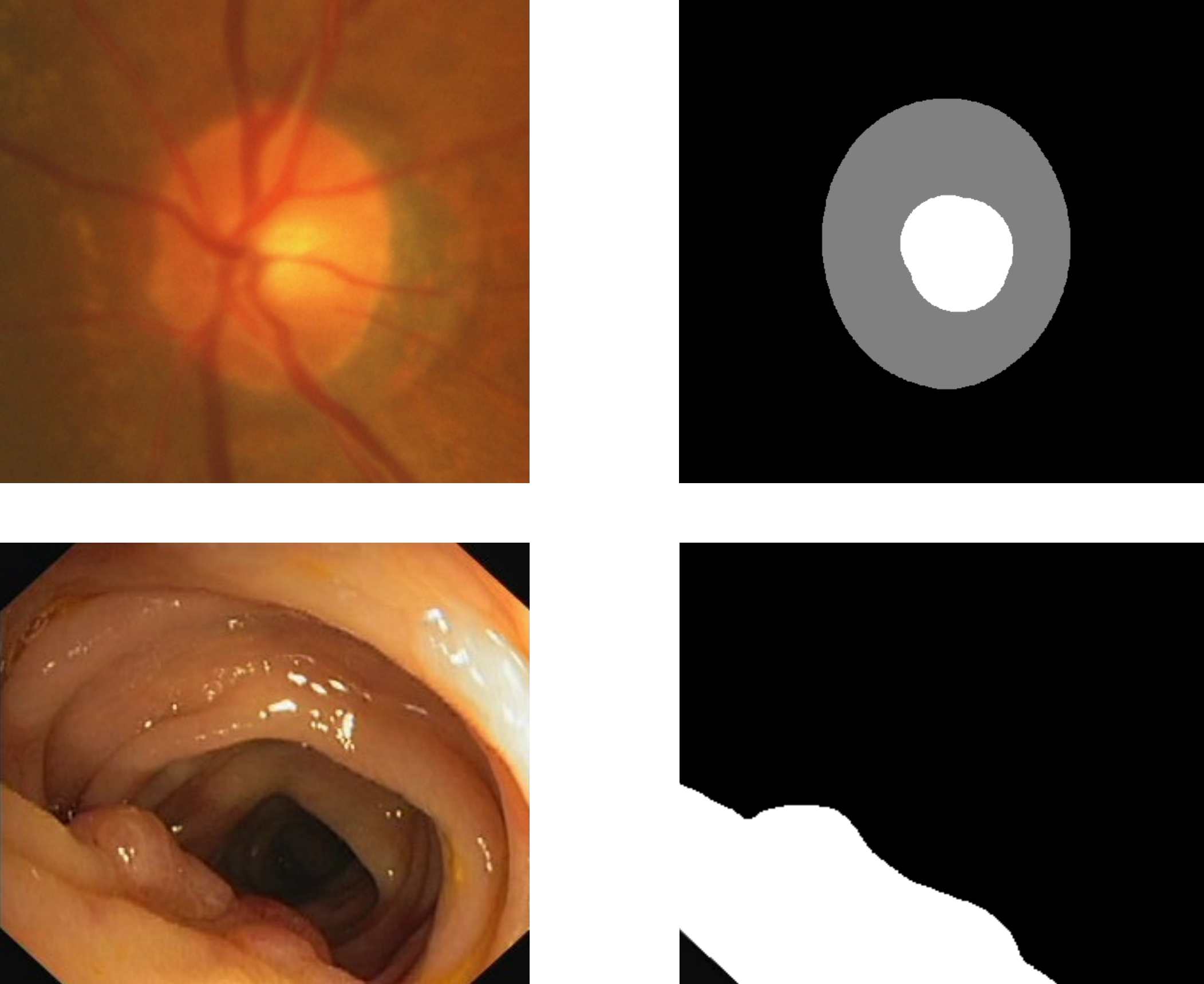}
  \captionof{figure}{Top: Optic disc/cup segmentation in fundus images into 3 classes: disc (grey), cup (white), and background (black). Bottom: Polyp segmentation in colonoscopy images into 2 classes: polyp (white) and background (black).}
  \label{fig:refuge-polyp}
\end{figure}
Three tasks were evaluated in our experiments:

\paragraph{REFUGE20: Optic Disc/Cup Segmentation in Fundus Images.}
This task does segmentation of the optic disc and cup in fundus images, which are 2D images of the rear of eyes (Fig. \ref{fig:refuge-polyp}). It is a subtask of the REFUGE  Challenge\footnote{\url{https://refuge.grand-challenge.org/Home2020/}} \cite{refuge}, MICCAI 2020. 1200 images were provided for training, and 400 for validation. We also used two extra datasets, Drishti-GS dataset \cite{drishti} and RIM-ONE v3 \cite{rim-one} when training all models. The \textbf{Disc}/\textbf{Cup} dice scores of validation images were obtained from the official evaluation server.

\paragraph{Polyp: Polyp Segmentation in Colonoscopy Images.}
Polyps are fleshy growths in the colon lining that may become cancerous. This task does polyp segmentation in colonoscopy images (Fig. \ref{fig:refuge-polyp}). Two image datasets \cite{pranet} were used: CVC612 (\textbf{CVC} in short; 612 images) and \textbf{Kvasir} (1000 images). Each was randomly split into 80\% training and 20\% validation, and the training images were merged into one set.

\paragraph{BraTS19: Tumor Segmentation in MRI Images.}
This task focuses on the segmentation of gliomas, a common brain tumor in MRI scans. It was a subtask of the BraTS'19 challenge\footnote{\url{https://www.med.upenn.edu/cbica/brats-2019/}}  \cite{brats-tmi,brats-sci-data}, MICCAI 2019. It involves four classes: the whole tumor (\textbf{WT}), the tumor core (\textbf{TC}), the enhancing tumor (\textbf{ET}) and background. Among them, the tumor core consists of the necrotic regions and non-enhancing tumors (red), as well as the enhancing tumor (yellow). 335 scans were provided for training, and 125 for validation. The dice scores of ET, WT and TC on the validation scans were obtained from the official evaluation server.

\subsection{Ablation Studies}
Two ablation studies were performed on REFUGE20 to compare: 1) the Squeeze-and-Expansion Transformer versus Multi-Head Transformer; and 2) the Learnable Sinusoidal Positional Encoding versus two schemes as well as not using PE.

All the settings were variants of the standard one, which used three layers of Squeeze-and-Expansion transformer with four modes ($N_m=4$), along with learnable sinusoidal positional encoding. Both ResNet-101 and EfficientNet-B4 were evaluated to reduce random effects from choices of the backbone. We only reported the cup dice scores, as the disc segmentation task was relatively easy, with dice scores only varying $\pm 0.005$ across most settings. 

\paragraph{Type of Transformer Layers.}
Table \ref{layer-scheme} shows that Squeeze-and-Expansion transformer outperformed the traditional multi-head transformers. Moreover, Both the squeeze attention block and the expansion attention block contributed to improved performance.

\begin{table}[h]
\begin{centering}\setlength\tabcolsep{3pt}
\begin{tabular}{|c|c|c|}
\hline 
Transformer Type & ResNet-101 & Eff-B4 \tabularnewline
\hline 
Cell-DETR ($N_h=4$) &  0.846 & 0.857 \tabularnewline
\hline 
Multi-Head ($N_h=4$) &  0.858 & 0.862 \tabularnewline
\hline 
No squeeze + Expansion ($N_m=4$) &  0.840 & \textbf{0.872} \tabularnewline
\hline 
Squeeze + Single-Mode &  0.859 & 0.868 \tabularnewline
\hline
\textbf{Squeeze + Expansion ($N_m=4$)} &  \textbf{0.862} & \textbf{0.872} \tabularnewline
\hline 
\end{tabular}
\caption{REFUGE'20 Fundus Optic Cup dice scores change with the type of transformer layers. Single-Mode implies No Expansion. Cell-DETR uses a multi-head transformer and discretely learned PE. $N_h$: number of attention heads in a MHA. $N_m$: number of modes in a Squeeze-and-Expansion transformer. }
\label{layer-scheme}
\par\end{centering}
\end{table}

\paragraph{Positional Encoding.}
Table \ref{pos-scheme} compares learnable sinusoidal positional encoding with the two mainstream PE schemes and no PE. Surprisingly, without PE, performance of Segtran only dropped 1\textasciitilde 2\%. A possible explanation is that the transformer may manage to extract positional information from the CNN backbone features \cite{cnn-pos}.
\begin{table}[h]
\begin{centering}
\begin{tabular}{|c|c|c|}
\hline 
Positional Encoding & ResNet-101 & Eff-B4 \tabularnewline
\hline 
None & 0.857  & 0.853 
\tabularnewline
\hline 
Discretely learned & 0.852 & 0.860 \tabularnewline
\hline 
Fixed Sinusoidal & 0.857  & 0.849 \tabularnewline
\hline
\textbf{Learnable Sinusoidal} &  \textbf{0.862} & \textbf{0.872} \tabularnewline
\hline 
\end{tabular}
\caption{REFUGE'20 Fundus Optic Cup dice scores change with the type of positional encoding (PE) schemes.}
\label{pos-scheme}
\par\end{centering}
\end{table}

\paragraph{Number of Transformer Layers.} 
Table \ref{num-layers} shows that as the number of transformer layers increased from 1 to 3, the performance improved gradually. However, one more layer caused performance drop, indicating possible overfitting.

\begin{table}[h]
\begin{centering}
\begin{tabular}{|c|c|c|}
\hline 
Number of layers & ResNet101 & Eff-B4 \tabularnewline
\hline 
1 & 0.856 & 0.854 \tabularnewline
\hline 
2 & \textbf{0.862} & 0.857 \tabularnewline
\hline
\textbf{3} & \textbf{0.862} &  \textbf{0.872} \tabularnewline
\hline
4 & 0.855 &  0.869 \tabularnewline
\hline 
\end{tabular}
\caption{REFUGE20 Optic Cup dice scores change with the number of transformer layers. Best performers with each backbone are highlighted.}
\label{num-layers}
\par\end{centering}
\end{table}

\subsection{Comparison with Baselines}
Ten methods were evaluated on the 2D segmentation tasks:
\begin{itemize}
    \item \textbf{U-Net} \cite{unet}: The implementation in a popular library \emph{Segmentation Models.PyTorch} (SMP) was used\footnote{\url{https://github.com/qubvel/segmentation_models.pytorch/}}. The pretrained ResNet-101 was chosen as the encoder. In addition, U-Net implemented in U-Net++ (below) was evaluated as training from scratch. 
    
    \item \textbf{U-Net++} \cite{unet++}: A popular PyTorch implementation\footnote{\url{https://github.com/4uiiurz1/pytorch-nested-unet}}. It does not provide options to use pretrained encoders, and thus was only trained from scratch.
    
    \item \textbf{U-Net3+} \cite{unet3+}: The official PyTorch implementation\footnote{\url{https://github.com/ZJUGiveLab/UNet-Version}}. It does not provide options to use pretrained encoders.
    
    \item \textbf{PraNet} \cite{pranet}: The official PyTorch implementation\footnote{\url{https://github.com/DengPingFan/PraNet}}. The pretrained Res2Net-50 \cite{res2net} was recommended to be used as the encoder.
    
    \item \textbf{DeepLabV3+} \cite{deeplabv3+}: A popular PyTorch implementation\footnote{\url{hhttps://github.com/VainF/DeepLabV3Plus-Pytorch}}, with a pretrained ResNet-101 as the encoder.
    
    \item \textbf{Attention based U-Nets} \cite{att-unet}: Attention U-Net (\textbf{AttU-Net}) and \textbf{AttR2U-Net} (a combination of AttU-Net and Recurrent Residual U-Net) were evaluated\footnote{\url{https://github.com/LeeJunHyun/Image_Segmentation}}. They learn to focus on important areas by computing element-wise attention weights (as opposed to the pairwise attention of transformers).
    
    \item \textbf{nnU-Net} \cite{nnunet}: nnU-Net generates a custom U-Net configuration for each dataset based on its statistics. It is primarily designed for 3D tasks, but can also handle 2D images after converting them to pseudo-3D. The original pipeline is time-consuming, and we extracted the generated U-Net configuration and instantiated it in our pipeline to do training and test.
    
    \item \textbf{Deformable U-Net} \cite{dunet}: Deformable U-Net (\textbf{DUNet}) uses deformable convolution in place of ordinary convolution. The official implementation
\footnote{\url{https://github.com/RanSuLab/DUNet-retinal-vessel-detection}} 
of DUNet was evaluated.

    \item \textbf{SETR} \cite{SETR}: SETR uses ViT as the encoder, and a few convolutional layers as the decoder. The SETR-PUP model in the official implementation\footnote{\url{https://github.com/fudan-zvg/SETR/}} was evaluated, by fine-tuning the pretrained ViT weights.
    
    \item \textbf{TransU-Net} \cite{transunet}: TransU-Net uses a hybrid of ResNet and ViT as the encoder, and a U-Net style decoder. The official implementation\footnote{\url{https://github.com/Beckschen/TransUNet}} was evaluated, by fine-tuning their pretrained weights.

    \item \textbf{Segtran}: Trained with either a pretrained ResNet-101 or EfficientNet-B4 as the backbone.
\end{itemize}
Three methods were evaluated on the 3D segmentation task:
\begin{itemize}
    \item \textbf{Extension of nnU-Net} \cite{nnu-net-ext}: An extension of the nnU-Net\footnote{\url{https://github.com/woodywff/brats_2019}} with two sampling strategies.
    \item \textbf{Bag of tricks (2nd place solution of the BraTS'19 challenge)} \cite{bag-tricks}: The winning entry used an ensemble of five models. For fairness, we quoted the best single-model results (``BL+warmup'').
    \item \textbf{Segtran-3D}: I3D \cite{i3d} was used as the backbone.
\end{itemize}

\subsection{Training Protocols}
All models were trained on a 24GB Titan RTX GPU with the AdamW optimizer. The learning rate for the three transformer-based models were 0.0002, and 0.001 for the other models. On REFUGE20, all models were trained with a batch size of 4 for 10,000 iterations (27 epochs); on Polyp, the total iterations were 14,000 (31 epochs). On BraTS19, Segtran was trained with a batch size of 4 for 8000 iterations.

The training loss was the average of the pixel-wise cross-entropy loss and the dice loss. 
Segtran used 3 transformer layers on 2D images, and 1 layer on 3D images to save RAM. The number of modes in each transformer layer was 4.

\subsection{Results}

\begin{table}[t]
\begin{centering}\setlength\tabcolsep{2pt}
\begin{tabular}{|m{3.3cm}|c|c|c|c|c|}
\hline 
& \multicolumn{2}{c|}{REFUGE20} & \multicolumn{2}{c|}{Polyp} & Avg.\tabularnewline
\cline{2-5} 
 & Cup & Disc & Kvasir & CVC & \tabularnewline
\hline 
U-Net & 0.730 & 0.946 & 0.787 & 0.771 & 0.809 \tabularnewline
\hline 
U-Net (R101) & 0.837 & 0.950 & 0.868 & 0.844 & 0.875 \tabularnewline
\hline 
U-Net++ & 0.781 & 0.940 & 0.753 & 0.740 & 0.804 \tabularnewline
\hline 
U-Net3+ & 0.819 & 0.943 & 0.708 & 0.680 & 0.788 \tabularnewline
\hline 
PraNet (res2net50) & 0.781 & 0.946 & 0.898 & 0.899 & 0.881 \tabularnewline
\hline 
DeepLabV3+ (R101) & 0.839 & 0.950 & 0.805 & 0.795 & 0.847 \tabularnewline
\hline 
AttU-Net & 0.846 & 0.952 & 0.744 & 0.749 & 0.823 \tabularnewline
\hline 
AttR2U-Net & 0.818 & 0.944 & 0.686 & 0.632 & 0.770 \tabularnewline
\hline 
DUNet & 0.826 & 0.945 & 0.748 & 0.754 & 0.818 \tabularnewline
\hline 
nnU-Net & 0.829 & 0.953 & 0.857 & 0.864 & 0.876 \tabularnewline\hline
SETR (ViT) & 0.859 & 0.952 & 0.894 & 0.916 & 0.905 \tabularnewline
\hline 
TransU-Net (R50+ViT) & 0.835 & 0.958 & 0.895 & 0.916 & 0.901 \tabularnewline
\hline 
Segtran (R101) & 0.862 & 0.956 & 0.888 & 0.929 & 0.909 \tabularnewline
\hline 
Segtran (eff-B4) & \textbf{0.872} & \textbf{0.961} & \textbf{0.903} & \textbf{0.931} & \textbf{0.917} \tabularnewline
\hline 
\end{tabular}
\caption{Dice scores on REFUGE20 and Polyp validation sets. R101: ResNet-101; R50: ResNet-50; eff-B4: EfficientNet-B4.}
\label{scores2d}
\par\end{centering}
\end{table}

\begin{table}
\begin{centering}
\begin{tabular}{|c|c|c|c|c|c|}
\hline 
\multirow{2}{*}{} & \multicolumn{4}{c|}{BraTS19}\tabularnewline
\cline{2-5}
 & ET & WT & TC & Avg. \tabularnewline
\hline 
Extension of nnU-Net & 0.737 & 0.894 & 0.807 & 0.813 \tabularnewline
\hline 
Bag of tricks & 0.729 & \textbf{0.904} & 0.802 & 0.812 \tabularnewline
\hline 
Segtran (i3d) & \textbf{0.740} & 0.895 & \textbf{0.817} &  \textbf{0.817} \tabularnewline
\hline 
\end{tabular}
\caption{Dice scores on BraTS19 validation set. Only single-model performance is reported.}
\label{scores3d}
\par\end{centering}
\end{table}

Tables \ref{scores2d} and \ref{scores3d} present the evaluation results on the 2D and 3D tasks, respectively. Overall, the three transformer based methods, i.e., SETR, TransU-Net and Segtran achieved best performance across all tasks. With ResNet-101 as the backbone, Segtran performed slightly better than SETR and TransU-Net. With EfficientNet-B4, Segtran exhibited greater advantages.

It is worth noting that, Segtran (eff-B4) was among the top 5 teams in the semifinal and final leaderboards of the REFUGE20 challenge. Among either REFUGE20 or BraTS19 challenge participants, although there were several methods that performed slightly better than Segtran, they usually employed ad-hoc tricks and designs \cite{refuge,nnu-net-ext,bag-tricks}. In contrast, Segtran achieved competitive performance 
with the same architecture and minimal hyperparameter tuning, free of domain-specific strategies.

\begin{figure}[h]
\centering
  \includegraphics[scale=0.25]{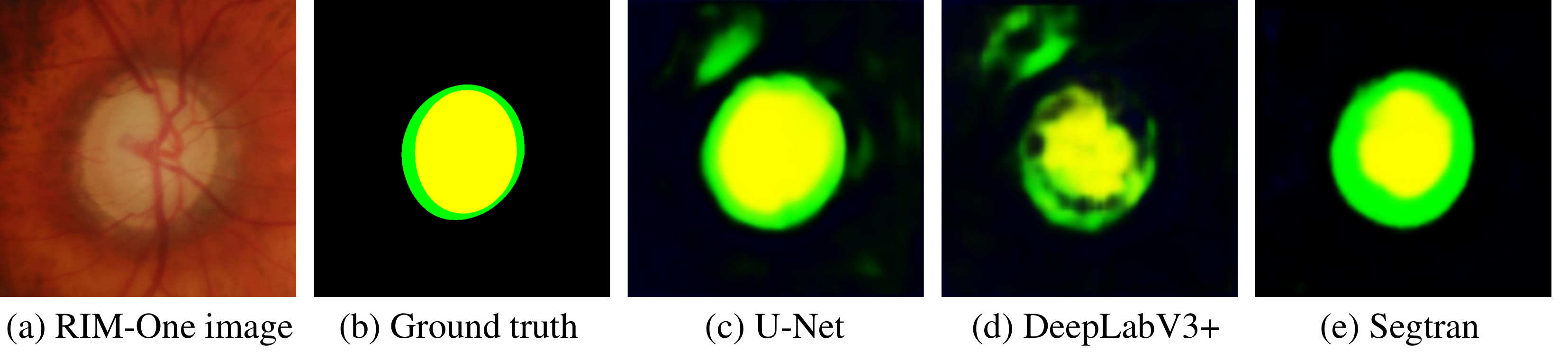}
  \captionof{figure}{Soft segmentation masks produced by different methods on a RIM-One image. The mask by Segtran has the fewest artifacts.}
  \label{fig:rim-one}
\end{figure}

\subsection{Cross-Domain Generalization}
To explore how well different methods generalize to new domains, we trained three representative methods, U-Net, DeepLabV3+ and Segtran on the 1200 training images of REFUGE20. All the methods used a pretrained ResNet-101 as the encoder/backbone. The trained models were evaluated on both the REFUGE20 training images and the RIM-One dataset \cite{rim-one}. As RIM-One images have drastically different characteristics from REFUGE20, all models suffered severe performance drop, as shown in Table \ref{generalization}. Nevertheless, Segtran had the least performance degradation, showing the best cross-domain generalization. Fig.\ref{fig:rim-one} shows a RIM-One image and the corresponding soft segmentation masks (before thresholding) produced by different methods. The mask produced by Segtran contains the fewest artifacts.

\begin{table}[h]
\begin{centering}
\begin{tabular}{|l|c|c|c|}
\hline 
 & REFUGE & RIM-One & Drop \tabularnewline
\hline 
U-Net & 0.862 & 0.680 & -0.182 \tabularnewline
\hline 
DeepLabV3+ & 0.846 & 0.653 & -0.193 \tabularnewline
\hline 
Segtran & \textbf{0.938} & \textbf{0.796} & \textbf{-0.142} \tabularnewline
\hline 
\end{tabular}
\caption{Generalization of three methods, measured by drop of Optic Cup dice scores from the REFUGE20 training images to a new test domain RIM-One. The smaller the drop is, the better. All used ResNet-101 as the encoder/backbone.}
\label{generalization}
\par\end{centering}
\end{table}

\subsection{Computational Efficiency}
Table \ref{flops} presents the number of parameters and FLOPs of a few representative methods. In general, transformer-based methods consume more computation and GPU RAM than conventional methods.

Our profiling showed that the number of parameters/FLOPs of Segtran are dominated by the output FPN, which vary drastically across different backbones. As the bottom-up FPNs we adopt are somewhat similar to EfficientDet \cite{efficientdet}, the model size/FLOPs are optimal when using EfficientNets. With ResNets as the backbone, Segtran has a significantly higher model size/FLOPs, and hence this choice of backbone is not recommended for efficiency-sensitive scenarios.

\begin{table}[h]
\begin{centering}
\begin{tabular}{|m{3.5cm}|c|c|}
\hline
& Params (M) & FLOPs (G) \tabularnewline \hline 
nnU-Net & 41.2 & 16.3 \tabularnewline \hline 
AttU-Net & 34.9 & 51.0  \tabularnewline \hline 
SETR (ViT) & 307.1 & 91.1  \tabularnewline \hline 
TransU-Net (R50+ViT) & 93.2 & 32.2  \tabularnewline \hline 
Segtran (R101) & 166.7 & 152.8 \tabularnewline \hline 
Segtran (eff-B4) & 93.1 & 71.3 \tabularnewline \hline 
\end{tabular}
\caption{Number of parameters / FLOPs on a 256x256 input image.} \label{flops}
\par\end{centering}
\end{table}

\subsection{Impact of Pretraining}
Models for medical image tasks usually benefit from initialization with weights pretrained on natural images (e.g. ImageNet \cite{imagenet}), as medical image datasets are typically small. To quantitatively study the impact of pretraining, Table \ref{pretraining} compares the performance of using pretrained weights vs. training from scratch of a few methods. Pretraining brought \textasciitilde2.5\% increase of average dice scores to the two transformer-based models, and 1\% to U-Net (ResNet-101).

\begin{table}[h]
\begin{centering}\setlength\tabcolsep{2pt}
\begin{tabular}{|m{3.3cm}|c|c|c|c|c|}
\hline 
& \multicolumn{2}{c|}{REFUGE20} & \multicolumn{2}{c|}{Polyp} & Avg.\tabularnewline
\cline{2-5} 
 & Cup & Disc & Kvasir & CVC & \tabularnewline
\hline 
U-Net (R101 scratch) & 0.827 & 0.953 & 0.847 & 0.835 & 0.865 \tabularnewline
\hline 
U-Net (R101 pretrain) & 0.837 & 0.950 & 0.868 & 0.844 & 0.875 \tabularnewline
\hline 
TransU-Net (R50+ViT scratch) & 0.817 & 0.943 & 0.869 & 0.872 & 0.875 \tabularnewline
\hline 
TransU-Net (R50+ViT pretrained) & 0.835 & \textbf{0.958} & \textbf{0.895} & 0.916 & 0.901 \tabularnewline
\hline 
Segtran (R101 scratch) & 0.852 & 0.939 & 0.858 & 0.851 & 0.875 \tabularnewline
\hline 
Segtran (R101 pretrain) & \textbf{0.862} & 0.956 & 0.888 & \textbf{0.929} & \textbf{0.909} \tabularnewline
\hline 
\end{tabular}
\caption{Impact of using pretrained encoder weights.} \label{pretraining}
\par
\end{centering}
\end{table}

\section{Conclusions}
In this work, we present Segtran, a transformer-based medical image segmentation framework. It leverages unlimited receptive fields of transformers to contextualize features. Moreover, the transformer is an improved Squeeze-and-Expansion transformer that better fits image tasks. Segtran sees both the global picture and fine details, lending itself good segmentation performance. On two 2D and one 3D medical image segmentation tasks, Segtran consistently outperformed existing methods, and generalizes well to new domains.

\section*{Acknowledgements}
We are grateful for the help and support of Wei Jing. This research is supported by A*STAR under its Career Development Award (Grant No. C210112016), and its Human-Robot Collaborative Al for Advanced Manufacturing and Engineering (AME) programme (Grant No. A18A2b0046).

\bibliographystyle{named}
{\small
\bibliography{segtran}}
\end{document}